\documentclass[natbib209,12pt,preprint]{aastex}

\newcommand{\ace}{$\alpha_{CE}$}

\newcommand{\msun}{$M_\odot$}
\newcommand{\lsun}{$L_\odot$}
\newcommand{\rsun}{$R_\odot$}

\newcommand{\kms}{km~s$^{-1}$}

\def \la{\mathrel{<\kern-1.0em\lower0.9ex\hbox{$\sim$}}}
\def \ga{\mathrel{>\kern-1.0em\lower0.9ex\hbox{$\sim$}}}

\shorttitle{PN paradigm}
\shortauthors{De Marco et al.}

\begin{document}

\title{Binary central stars of PN discovered through photometric variability. 
I. What we know and what we would like to find out}

\author{Orsola De~Marco$^1$, Todd C.~Hillwig$^2$, \& A.J.~Smith$^{3,4}$}

\altaffiltext{1}{Dept. of Astrophysics, American Museum of Natural History, New York, NY 10024}
\altaffiltext{2}{Dept. of Physics and Astronomy, Valparaiso University, Valparaiso, IN 46383}
\altaffiltext{3}{Dept. of Physics, University of California at Davis, Davis, CA 95616} 
\altaffiltext{4}{REU student at the Dept. of Astrophysics, American Museum of Natural History}

\begin{abstract}
Shaping axi-symmetric planetary nebulae is easier if a companion interacts with a primary at the top of the asymptotic giant branch. To determine the impact of binarity on planetary nebula formation and shaping, we need to determine the central star of planetary nebula binary fraction and period distribution. The short-period binary fraction has been known to be 10-15\% from a  survey of $\sim$100 central stars for photometric variability indicative of irradiation effects, ellipsoidal variability or eclipses. This survey technique is known to be biased against binaries with long periods and this fact is used to explain why the periods of all the binaries discovered by this survey are smaller than 3 days. In this paper we assess the status of knowledge of binary central stars discovered because of irradiation effects. We determine that, for average parameters, this technique should be biased against periods longer than 1-2 weeks, so it is surprising that no binaries were found with periods longer than 3 days. Even more puzzling is the fact that 9 out of 12 of the irradiated binaries, have periods smaller than {\it one day}, a fact that is starkly at odds with post-common envelope predictions. We suggest that either all common envelope models tend to overestimate post-common envelope periods or that this binary survey might have suffered from additional, unquantified biases. If the latter hypothesis is true, the currently-known short-period binary fraction is put in serious doubt. We also introduce a new survey for binary-related variability, which will enable us to better quantify biases and determine an independent value for the short period binary fraction.

\end{abstract}


\section{Introduction}
\label{sec:introduction}

The importance of binarity in the production and shaping of planetary nebulae (PNe) has been actively debated for the last three decades with a minority of authors arguing for a more prominent role of binary interactions, such as common envelopes \citep[CE; e.g.,][]{Bond2000,Soker1997}. In the single star scenario, magnetic fields and stellar rotation have been deemed sufficient to achieve most of the observed shapes \citep[e.g.,][]{GarciaSegura1999}. However, recently we have appreciated that a single asymptotic giant branch (AGB) star cannot sustain large scale magnetic fields long enough for it to act on the mass-loss geometry \citep{Nordhaus2007,Soker2006}. A rotating AGB star can produce a large scale magnetic field, but this field will drain angular momentum from the star, slow down the rotation that sustains it, and therefore switch itself off. To maintain the magnetic field, an accessible angular momentum source, most likely in the form of a companion is needed. 

The central star of PN close binary fraction has been determined by a survey of $\sim$100 central stars for photometric variability indicative of irradiation effects, ellipsoidal variation or eclipses. The survey was started in the 1970s and was carried out primarily by H.E. Bond, A.D. Grauer and R. Ciardullo. Its results have been presented on several occasions \citep{Bond1985,Bond1987,Bond1992,Bond1995,Ciardullo1996,Bond2000}. We shall therefore call this the BGC survey.
From this survey a binary fraction of 10-15\% was determined (\citealt{Bond2000}, see also \citealt{DeMarco2006}). This number is not high enough to be consistent with a large fraction of PNe having been shaped by binary interactions, even supposing that some binary interactions end in mergers and others leave binary central stars with periods too long to be easily detected. However, all binaries detected by the BGC survey have periods smaller than 3 days, with most of them having periods smaller than one day. This has been explained by arguing that longer periods (hence larger separations) lead to much reduced irradiation effects, ellipsoidal variation and the chance of eclipses. Hence the binary fraction of 10-15\% has long been considered biased towards {\it very} short periods and has been considered a lower limit. 

Yet the exact bias of the BGC survey is hard to quantify since it is not clear from the literature what detections limits were in place. These limits depended on the specific observing run (the survey took place over three decades with many observing runs at different telescopes and with different conditions), as well as on the variability of the targets caused by other effects such as pulsations \citep[see for instance][]{Ciardullo1996}, or winds. We should also take into consideration that, the relentless work by H. Bond and collaborators was, at least initially, biased towards central stars suspected of variability by other workers \citep[see Section~II of][]{Grauer1983}, and this could have introduced a bias towards binarity.

All of the central star binaries detected by the BGC survey are believed to be post-CE binaries, where the primary and secondary spiraled in towards each other when the primary's expanding envelope engulfed the secondary \citep{Paczynski1976}. The post-CE period distributions predicted by population synthesis models \citep[e.g.,][]{Yungelson1993,Han1995} show that the binary frequency is an increasing function of period (for $P \la 1$ day), for the entire range of plausible assumptions. The BGC survey shows a {\it decrease} in the number of binaries for periods longer than {\it only} $\sim$0.4~days. Unless this decrease is justified by a detection bias against periods longer than that, the observed period distribution is at odds with predictions.


An alternative method to search for close binaries is by the radial velocity technique, capable in principle of finding binaries with longer periods. This technique has unfortunately been plagued by the pervasive and dramatic wind and pulsation-induced spectral variability\footnote{While spectral variability induced by variable clumpy winds is well documented \citep{Patriarchi1995,Patriarchi1997}, spectral variability due to pulsations is not as well understood in hot, H-rich central stars \citep{Handler2003,DeMarco2008a}.}, which make the stellar spectra exhibit radial velocity shifts thus introducing noise and making it difficult to detect periods \citep{Mendez1991,Sorensen2004,DeMarco2004,Afsar2005,DeMarco2008a}. Efforts are under-way to carry our radial velocity surveys of fainter central stars, less likely to have winds and pulsations. For now, however, the central star of PN binary fraction remains undetermined.

A PN population dominated ($\sim$70\%) by binaries that either went though a CE or some other form of strong interaction while the primary was on the AGB (orbital periods of the central star binary smaller than $\sim$10\,000 days), is consistent with what we know of main sequence binarity, stellar and galactic evolution, in the sense that there are enough binaries on the main sequence with the appropriate characteristics to account for most of today's Galacitc PN population (\citealt{DeMarco2005}, \citealt{Moe2006}, De~Marco et al., in prep)\footnote{In the binary scenario there would be a population of single central stars of under-luminous or invisible PN
\citep{Soker2005}.}. This would give us reason to think that indeed many close binary PN central stars  are yet to be found\footnote{At a recent meeting (Asymmetric PN IV, June 2007, http://www.iac.es/proyecto/apn4/) 120 participants, about 30-50\% of the active PN community, has agreed that this field is facing a paradigm problem and a working group (PlaN-\"B, for Planetary Nebula Binaries; www.wiyn.org/planb/) has been formed to test the binary hypothesis observationally.}.

In this paper we take a closer look at the known PN binaries discovered because of photometric variability, with the intention of determining whether a bias toward very short periods is evident from the sample. In so doing we review the properties of these binary systems revealing where more information is needed. This review is also fundamental when designing a new photometric survey whose biases might be easier to quantify. 

In Section~\ref{sec:knownbinaries} we summarize our current knowledge of close binary central stars known because of photometric variability (while details of each binary are given in the Appendix). 
In Section~\ref{sec:expectations} we discuss the theoretiacally-expected amplitude of the irradiation variability as a function of several binary and stellar parameters. This allows us to understand better the observed variability amplitude. In Section~\ref{sec:PvA} we discuss the amplitude of the variation as a function of orbital period and other parameters and use these diagrams as a platform to understand these binaries as a group. In Section~\ref{sec:thesurvey} we describe the characteristics of our new survey and how we avoid some of the caveats present in past surveys. Finally, we conclude in Section~\ref{sec:conclusions}.  

\section{The known, photometrically-variable central stars}
\label{sec:knownbinaries}

By 1985 \citep{Bond1985} four binaries were known as a result of the BGC survey. The number had increased to six by 1987 \citep{Bond1987} and to 11 by 1995 \citep{Bond1995}. In 2000 \citep{Bond2000}, we knew of 12 binaries from that survey. In this paper we list 12 binaries detected by photometric variability due to irradiation of the cool by the hot companion (though we have taken out SuWt~2 from the list of \citet{Bond2000}, since its variability is due to eclipses of two A-type stars (see Sec.~\ref{sssec:suwt2}) and included instead a new irradiated binary, NGC~6337 \citep{Hillwig2006}). One of these 12 binaries might have variability due to ellipsoidal variation and not to irradiation (A~41; Sec.~\ref{sssec:a41}). We further list 5 objects all of which have periodic photometric variability due to other causes (like ellipsoidal variability or dust) and one which we feel cannot be confirmed at this time (Hb~12; Sec.~\ref{sssec:hb12}).
In Appendix~A we review the details of the observations and models of each of the systems.  

A hot star can irradiate and heat one hemisphere of a close cool companion. Light variations are produced by the phase-dependent aspect of that side of the secondary which is heated and hence brighter. If the orbital inclination is close to 90 degrees, partial or total eclipses can be detected as an additional light variability with dips coinciding with minima of the irradiation light-curve. Finally, if one or both of the companion stars fill a significant fraction of their respective Roche lobes, the system appears brighter at the two quadratures than it does at the two conjunctions. This can introduce an additional modulation of the light-curve with a period that is half that of the orbital period.

In Table~\ref{tab:BCS}  we list the 12 known binaries whose light variation is thought to be due to an irradiation effect. Their PN and central star names (Columns 1 and 2) are followed by their $V$ brightness (determined by different authors as described in the text, where we list the minimum light brightness if known - if no description of the absolute brightness determination is found in the text then the brightness is that reported by the {\it Simbad} database (see sources within) and should be seen as an approximation of the system's brightness). The binary class is listed in Column 4, where ``I" stands for irradiation-induced variability, ``S1" and ``S2" stand for single- or double-lined spectroscopic binaries, ``Ec" stands for eclipsing binary, and ``El" stands for ellipsoidal variability. A question mark indicates that the nature of the variability is only presumed to be that indicated, but confirmation awaits. In Columns 5 and 6 the period and amplitude of the variability are listed, while in Column 7 we list the filter used. Finally, in Column 8 we state whether an emission line spectrum due to reprocessing of primary radiation by the heated side of the secondary was detected. All references can be found in the the Appendix (Sections \ref{sssec:beuma} to \ref{sssec:ngc6026}).

In Table~\ref{tab:stellarparameters} we list the stellar and system parameters derived in the literature by modeling of the spectra, light and radial velocity curves as well as other considerations. Parameters in italics are assumed, rather than derived by the light and radial velocoty curve analyses. Error bars are often formal errors and do not reflect other uncertainties, such as for instance the existence of very different models that fit the data equally well. However, if families of models that fit the data equally well do exist, we indicate that in the text and tables. The spectral types of the secondaries are listed in three columns: those derived from a spectrum of the night (non irradiated) side of the secondary,  those derived by assuming that the value of the mass is that of a main sequence star and those derived by assuming that the value of the radius is that of a main sequence star \citep{Cox2000}. The comparison of these three spectral types can aid in determining the status of the secondary star in the system. 

In Table~\ref{tab:stellarparameters2} we list parameters from the photometrically variable binaries whose variability is most likely due to causes other than irradiation, namely ellipsoidal variation or dust. We also list the central star of the PN Hb~12, which has been claimed to be an eclipsing binary, but whose light-curve, despite being seemingly periodic, is not straight-forwardly interpretable as due to eclipses. This object needs more observations (Sec.~\ref{sssec:hb12}).

\section{Theoretical relationships between amplitude of the irradiation-induced variability and stellar and system parameters}
\label{sec:expectations}

In order to take a critical look at the known irradiated binaries, we have run a Wilson-Devinney code \citep{Wilson1971,Wilson1990} for a set of typical stellar and system parameters and then determined the amplitude of the irradiation effect as we changed each of the parameters separately. 

We selected hypothetical system parameters (Table~\ref{tab:hypothetical}), similar to those of the real systems discussed in Section~\ref{sec:knownbinaries} (Table~\ref{tab:stellarparameters}). 
The effective temperature and radius of the primary were chosen so as to produce the bolometric luminosity of a 0.60~\msun\ central star (5030~\lsun) as determined by the hydrogen-burning tracks of \citet{Vassiliadis1994}. The secondary was chosen to have the mass of a main sequence star of spectral type M2V, but with a radius too large for that spectral type (0.6~\rsun, corresponding to a spectral type of M1V), in line with what we have observed in most of the known systems (Table~\ref{tab:stellarparameters}). The secondary effective temperature of 3500-K is approximately that of an M2V star \citep{Cox2000}. The reflection efficiencies of primary and secondary were 1.0 and 0.5, respectively (standard choices for high and low temperature stars).

Using this system as template we then varied, one at a time, parameters that play a role in the irradiation amplitude. Varying one parameter at a time can be unphysical. For instance as separation increases, the smaller irradiation might result in a smaller secondary radius (assuming, as we will do in Sec.~\ref{sec:conclusions}, that heating the atmosphere expands the star). For this exercise, however, we have concentrated on looking at individual parameters to understand in a quantitative, yet approximate manner, the complex interplay of variables in this type of model and shed further light on the systems that were analyzed in this way. 

\begin{figure}
\vspace{17cm}
\includegraphics{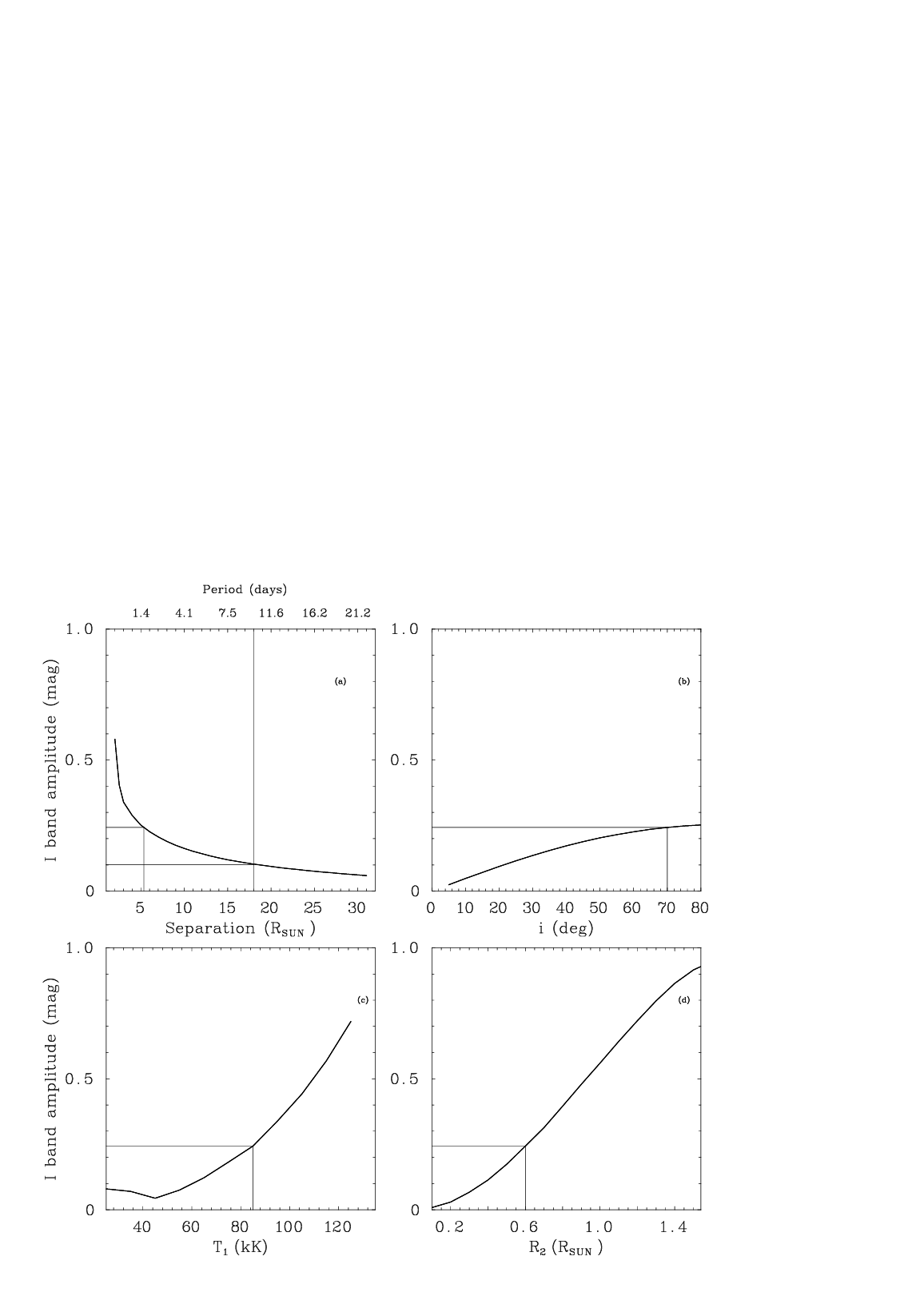}
\caption{(a) The amplitude of the irradiation effect (thick line) as a function of system's separation (bottom axis) or period (top axis), for the system parameters listed in Table~\ref{tab:hypothetical}. (b) The same as for plot (a) but as a function of orbital inclination. (c) The same as for plot (a) but as a function of primary effective temperature. (d) The same as plot (a) but as a function of secondary radius. The thin vertical and horizontal lines mark the irradiation amplitude for the model parameters listed in Table~\ref{tab:hypothetical}. The thicker grey lines in panel (a) marks the period/separation longward of which the hypothetical system might not be easily detectable (amplitude $<$ 0.1~mag).}
\label{fig:model}
\end{figure}

In Fig.~\ref{fig:model} (a) and (b) we present the light variability in the $I$ band as a function of orbital separation and orbital inclination, respectively. In Fig.~\ref{fig:model} (c) we present the amplitude as a function of primary effective temperature (where also the primary radius was varied so as to simulate a post-AGB star evolving to higher temperatures along a 0.6-\msun\ track). In Fig.~\ref{fig:model} (d) we present the amplitude as a function of secondary radius (i.e., the size of the irradiated surface). 

The relationships in panels (a), (b) and (d) of Fig.~\ref{fig:model} are easily explained by geometric effects. The amplitude vs. temperature relation (Fig.~\ref{fig:model} (c)) is slightly more complex because luminosity plays a role, where, for decreasing primary's radius, the primary's luminosity at first remains constant and then it decreases\footnote{We should also point out that a higher primary's luminosity does not necessarily result in a larger variability amplitude. This is due to  the fact that if the efficiency of the irradiation effect is less than one, the light of the brighter primary dominates over the brighter irradiation effect.  This means that the irradiation effect becomes a smaller ratio of the total light and the magnitude difference (or amplitude) actually decreases.}. The {\it increase} in amplitude for decreasing primary effective temperature, for temperature smaller than $\sim$45\,000~K is due to the fact that the primary starts to fill its Roche lobe and ellipsoidal variability is added to the irradiation effect (which, by itself, results in {\it decreasing} amplitude for decreasing primary effective temperature). 

\begin{figure}
\vspace{17cm}
\includegraphics{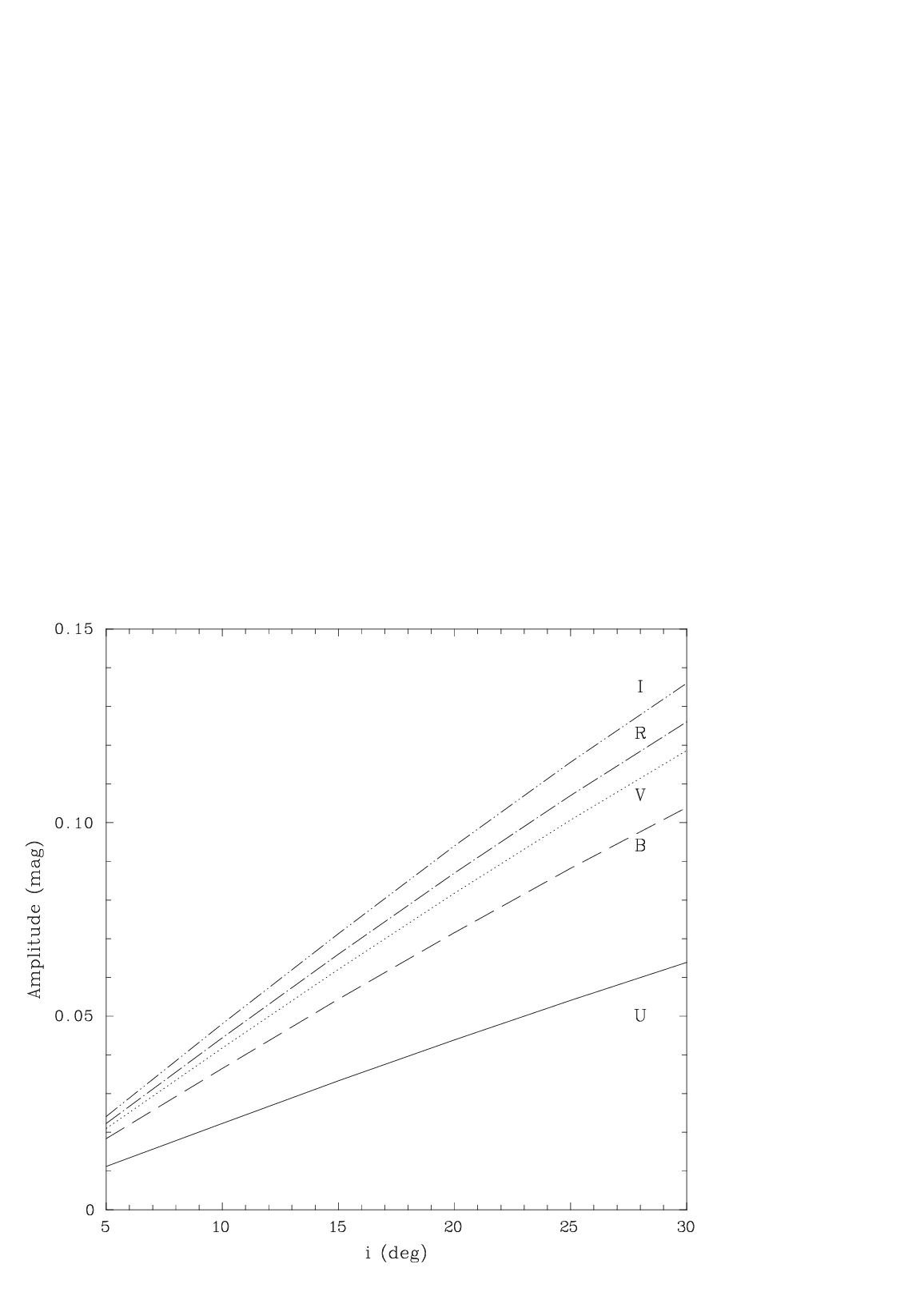}
\caption{The amplitude of the irradiation effect as a function of system's inclination, for 5 bands.}
\label{fig:BandComp}
\end{figure}

In Fig.~\ref{fig:BandComp} we show the same as in Fig.~\ref{fig:model} (b; in the range $i$=5-30~deg), but for different spectral bands ($U, B, V, R$ and $I$). It is interesting to note that while the $U$ and $B$ light variability amplitudes differ by approximately a factor of 1.6, there is very little difference (only 10-20\%) between the amplitudes in the $B, V, R$ and $I$ bands. These differences however, are not absolute and are themselves a function of, primarily, the total light contribution and color of the hot primary relative to the change in temperature and luminosity between irradiated and non-irradiated hemispheres of the companion. As an example of this complex behavior we point out the almost identical variability in all bands exhibited by HFG~1 (Table~\ref{tab:BCS} and Sec.~\ref{sssec:HFG1}).  

From this exercise we can also try to determine the bias of a survey like that of BGC. A system with an orbital separation of 18~\rsun, an inclination $\sim$70~deg and a primary temperature $\sim$85\,000~K, would have an $I$ band amplitude of 0.1~mag. If we assume that amplitudes smaller than 0.1~mag would not have been readily detected by the BGC survey (due to the less than optimal cadence, and to the addition of pulsational variability and stochastic variability due to winds), then our hypothetical system could not be observed for orbital separations $\ga$18~\rsun\ or periods $\ga$10~days. This hypothetical system has average parameters. A system with less (more) favorable parameters would become undetectable at a separation shorter (longer) than 18~\rsun. For example, an inclination of 30~deg, a primary temperature of 60\,000~K or a secondary radius of 0.4~\rsun, would each result in a reduction of the maximum separation at which detection of the variability can occur to about 7~\rsun\ or periods of 3-4 days. In the same way, a secondary radius of 0.8~\msun\ or an increase in theprimary temperature to 100\,000~K would allow us to detect the system beyond a separation of 30~\rsun\ or a period of $\sim$20~days\footnote{We note that if the secondary reflection efficiency (albedo) were higher (a lower value is unlikely), the reflection amplitude would be larger and the maximum period at which an irradiated binary would be detected would be longer. Hence the period calculated here as the period above which detection is unlikely is a conservative lower limit, emphasizing that it is highly unlikely that the observed maximum period of $\sim$3 days is an actual bias of the BGC survey.}. This theoretical information will later be combined with the information provided by the observations, in particular with the period distribution of the known systems, to draw a conclusion as to the likely bias of the BGC survey.  

\section{The observed period distribution and variability amplitudes\\ of the known binaries}
\label{sec:PvA}

\begin{figure}
\vspace{15cm}
\includegraphics{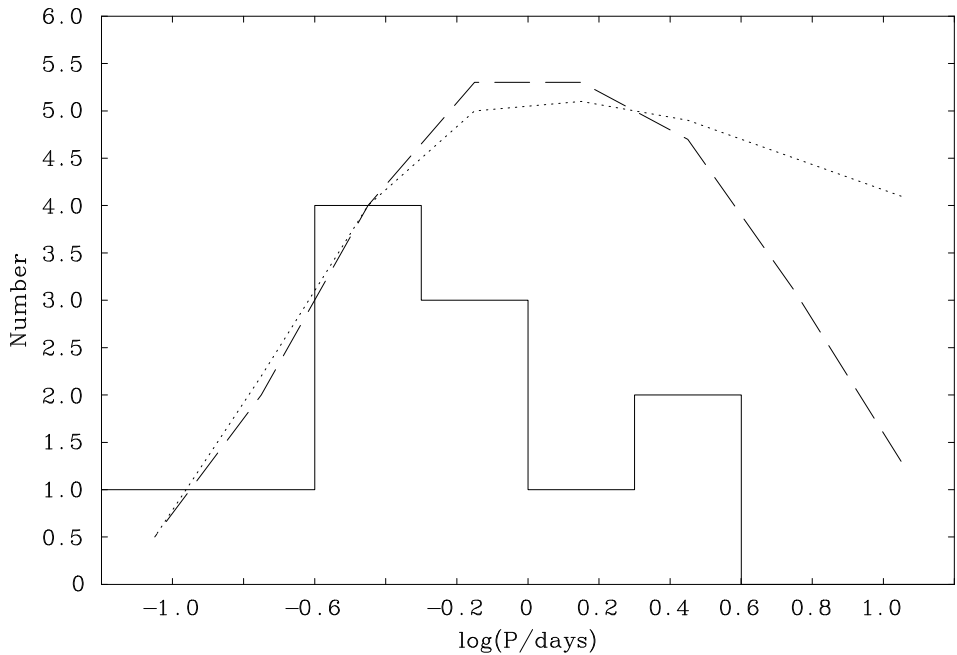}
\caption{Solid line: Period histogram for the observed irradiated binaries (Table~\ref{tab:BCS}). Dashed line: predicted period histogram from \citet{Han1995} for post-CE binaries with a main sequence companion where the ejection of the envelope was relatively inefficient. Dotted line: predicted period histogram from \citet{Han1995} for post-CE binaries with a main sequence companion where the ejection of the envelope was relatively efficient. The two theoretical distributions were scaled so as to have the same absolute value as the observations in the bin centered at log(P/days)=--0.45. As a result, each can be compared to the observations although they cannot be compared to each other (see text).}
\label{fig:Phist}
\end{figure}

In Fig.~\ref{fig:Phist} we show a histogram of the periods of the known irradiated binaries. First of all there is a disproportionate number of systems with period less than a day compared to longer than a day. Second, there are no systems with periods longer than 3 days. This could be due to (i) an intrinsically larger population of central star binaries with periods smaller than a day, or (ii) to a strong observational bias against periods longer than a few days already decreasing the number of detections for systems with periods longer than half a day. 

In Fig.~\ref{fig:Phist} we over-plot the post-CE period distribution predictions of \citet{Han1995} for an inefficient envelope ejection (low values of their parameters \ace\ and/or $\alpha_{th}$, which result in shorter post-CE periods) or an efficient one (high values of their parameters \ace\ and/or $\alpha_{th}$, which result in longer post-CE periods). Similar theoretical period distributions could be obtained from papers by \citet{Yungelson1993}, \citet{Han1995}, \citet{deKool1992} and others. Both distributions were normalized so as to have the same value as the observations in the bin $\log (P/$day$) = - 0.45$. This allows the two theoretical distributions to be compared to the data, but not to each other (for a comparison of the two theoretical distributions the reader is referred to Fig.4c of \citet{Han1995}, where it is evident that for the range of periods of interest to us, the efficient ejection predicts fewer objects). This comparison is used as an example to show that, irrespective of the model assumptions, one expects more systems with $P \ga 0.5$~day than observed. The observed period histogram of Fig.~\ref{fig:Phist} is definitely at odds with predictions. {\it If the observed period distribution were to be representative of reality (and not due to a bias -- see below), then the BGC survey would effectively demonstrate a problem with all our predictions for post-CE systems, with repercussion in our understanding of the periods of all post-CE classes such as CVs.}  

On the other hand, our theoretical study (Sec.~\ref{sec:expectations}) indicates that a strong observational bias capable of preventing detection of systems with periods larger than a couple of days and capable of systematically preventing detections of systems with P$\ga$3~days is also unlikely. There should be in fact plenty of binary systems with parameters that would allow them to be detected out to orbital separations of $\sim$18~\rsun\ (or periods of $\sim$10 days). Of course, considering the very small number of observed systems, one has to be careful in drawing this conclusion.

Another explanation could be that objects with very short periods have statistically larger photometric variability and have therefore been noticed preferentially in serendipitous discoveries and have then made their way into the BGC survey, unrealistically populating the very short period bin (see discussion of this pro-short period bias in the Introduction). If this is true, then it is possible that the central star binary fraction for periods shorter than a few days is actually smaller than the 10-15\% determined by the BGC survey.

\begin{figure}
\vspace{15cm}
\includegraphics{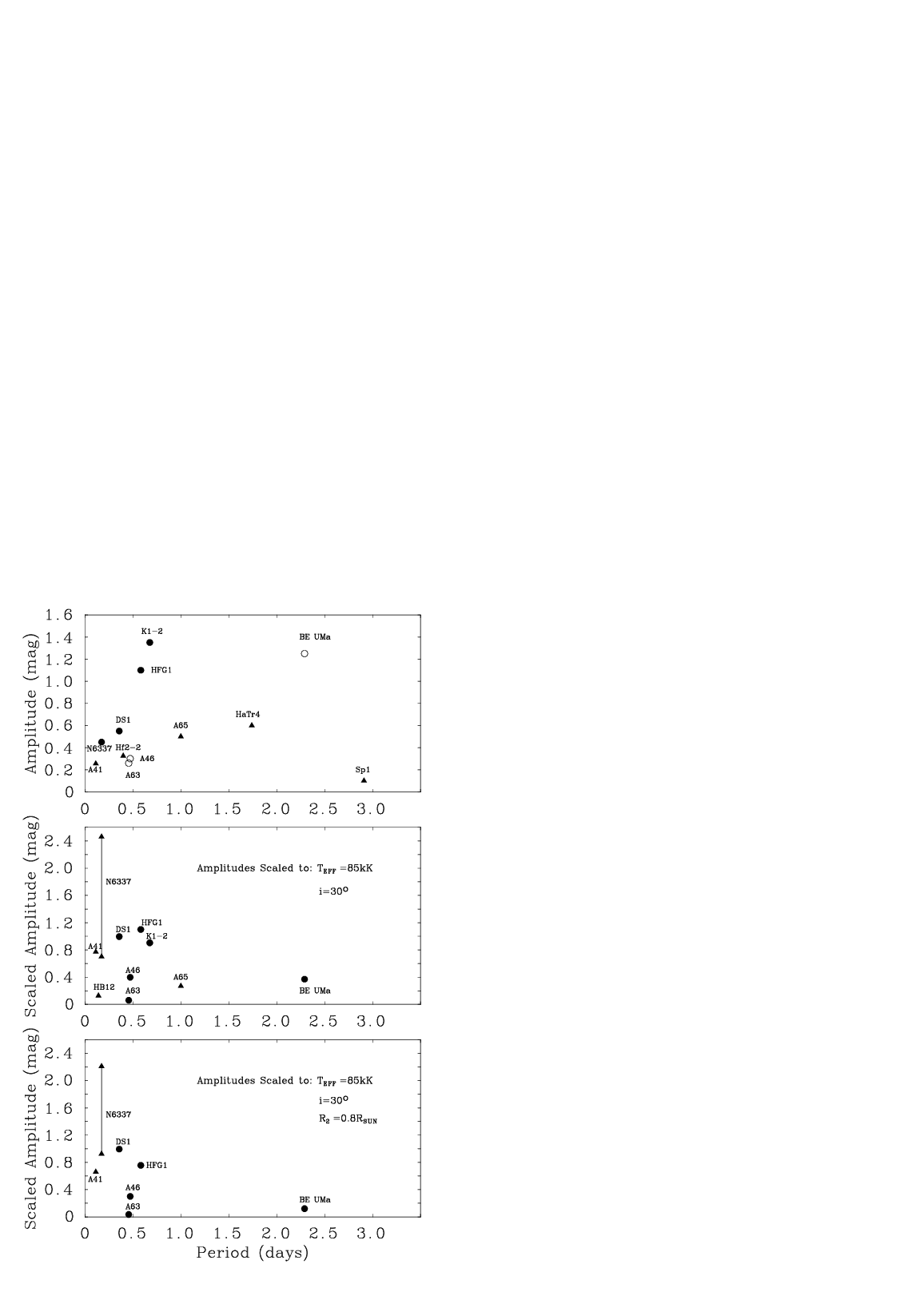}
\caption{Top panel: Period vs. amplitude for irradiated central stars of PN. Unfilled symbols are for the eclipsing binaries. Triangular symbols are for systems that need to be confirmed as irradiated binaries, while circular symbols are for confirmed systems. Middle panel: the variability amplitudes for all systems for which the primary effective temperature and the inclinationare known are scaled to an effective temperature of 85\,000~K and an inclination of 30~deg, using factors derived from Fig.~\ref{fig:model}. Bottom panel: the same as for the middle panel but where the amplitudes were further scaled to a secondary radius of 0.8~\rsun.}
\label{fig:periodvsamplitude}
\end{figure}

In an effort to further determine whether the available data gives clues as to the bias of the BGC
survey, we present, 
in Fig.~\ref{fig:periodvsamplitude} (top panel), a plot of the amplitudes of the irradiation effect in the $V$ band (or $B$ when no $V$ band information is available; these two bands are likely to have only a small inherent difference -- see Fig.~\ref{fig:BandComp} and Table~\ref{tab:BCS}), versus the period of the system. Since the total systems' masses for the central star binaries do not vary by much, the period is a reasonable proxy for the separation of the hot and cool components. Triangular symbols indicate systems for which more observations are needed and that should therefore be treated with caution. Considering the paucity of points on this diagram one should not be tempted to draw conclusions. However we here use this diagram as a reference point for an assessment of the situation. This exercise can help plan a new survey (Sec.~\ref{sec:thesurvey}), which will populate it with more systems and strengthen the statistics.


Theoretically, the amplitude of the variability does not only anti-correlate with separation, but also strongly correlates with primary temperature and secondary radius (and to a lesser extent with orbital inclination; Fig.~\ref{fig:model}). However, irrespective of the dependence of amplitude on several parameters, if hundreds of post-CE central stars were to populate the diagram in Fig.~\ref{fig:periodvsamplitude} (top panel), we would expect to see amplitudes decrease and the number of systems to dwindle, for increasing orbital separation and vanishes to zero for separations larger than a certain limit. In other words, we would expect, for a sufficient number of objects, that the separation bias of the survey would be revealed in a diagram like the one of Fig.~\ref{fig:periodvsamplitude} (top panel). 

Since we do not have a sufficient number of binaries and since several parameters play a role in the  variability amplitude, the decrease in amplitudes and number of systems as a function of separation is not readily visible. We therefore scaled the amplitudes to account for the effects of primary temperature and inclination (two of the parameters that play a role in the variability), using the relations determined by our models (Sec.~\ref{sec:expectations}). We scaled all amplitudes to a primary effective temperature of $T_1 = 85\,000$~K and a system inclination of 30~degrees.  When these scaled amplitudes are plotted (Fig.~\ref{fig:periodvsamplitude}; middle panel) the diagram shows an approximate anti-correlation of amplitude and period, as expected. This anti-correlation becomes more pronounced if we eliminate triangular symbols, which have highly uncertain values and if we chose, for NGC~6337, the low secondary temperature/low inclination  model (Table~\ref{tab:stellarparameters}), which has a larger scaled amplitude.

An additional parameter that greatly  affects amplitude of the photometric variability is the secondary's radius (Fig.~\ref{fig:model} (d)). For the few systems for which this information was available we scaled their variability amplitudes to $R_2 = 0.4$~\rsun. The correlation does improve somewhat (\ref{fig:periodvsamplitude}; bottom panel) with the loss of some of the systems that contributed to the scatter. A~41 remains anomalous in its behavior, as we will discuss in Sec.~\ref{sec:conclusions}, this might not be an irradiation binary after all.

First of all, this exercise reassures us that the relationship between amplitude and other parameters expected from the model (Sec.~\ref{sec:expectations}) are approximately in line with the data. It also shows that, given the set of chosen parameters, systems with period $\sim$3 days should indeed be easily detectable, as predicted from theory. If less favorable parameters are adopted, a system with a period of three days may fall below the detection threshold, but for every such system there should be a system with  more favorable parameters that would be detected. This, together with the unexplained decline of the number of systems for $P\ga 0.5$~day, leaves us without a reasonable explanation for the period distribution of the the BGC survey, and makes us question the meaning of the currently known binary fraction. 
  
 
\section{The new photometric variability survey for close binary central stars}
\label{sec:thesurvey}

In light of the considerations put forward in the previous sections we have initiated a new photometric variability survey of central stars of PN to carry out an independent determination of the central star close binary fraction. A new survey is also needed to increase the binary sample and allow us to improve the statistics of the binary central stars as a group. 

The new survey sample selection aims to minimize the variability due to effects other than binarity, that may compromise the detection of low amplitude periodic variability. H. Bond and collaborators (priv. comm.) reported that several central stars were variable. For some, this variability could be ascribed to pulsations \citep{Ciardullo1996}, while for others the variability remained unexplained and could possibly be due to winds. The telescope time and scheduling then available prevented them from fully characterizing those systems. Making use of the many observations already in hand (Bond priv. comm.) seems at first appealing. However, by doing so there would be no control on the sample selection, important when determining biases. The BGC survey was also carried out with several telescopes over 20 years and much of the key information that would allow us to determine biases such as weather conditions, cannot be trivially retrieved.

We have therefore set out to construct a new sample relatively free from winds and pulsations. This can be achieved by selecting only intrinsically faint central stars of old PNe. By selecting PN larger than a certain angular diameter {\it and} whose central stars are fainter than a certain apparent magnitude, one effectively selects intrinsically faint central stars\footnote{The exact limits vary somewhat depending on the telescope used for the survey (by necessity we have used and plan to use a range of telescopes with apertures between 1 and 3 meters, e.g., the SARA and 2.1~m and NOAO telescopes) as well as on the number of targets available for a given location and for a given season.}. The reason for this, is that PN that are large because they are close tend to have brighter central stars. This selection technique might have a slight bias against the most massive central stars because they fade fast and always remain in the middle of relatively small PN or have already faded beyond the detection limit of our telescopes. Another bias might be that spherical PNe, likely deriving from single stars, seem to be more frequent among fainter PNe \citep{Jacoby2008}. A target list populated by faint PN might therefore be biased against binarity. What matters is that for every observing session the exact selection applied to that target list, as well as the exact conditions and an estimate of the detection limits should be recorded. In this way the biases of each observing session can be combined together to assess the overall bias of a given sample.  

\citet{Hillwig2004} and \citet{Hillwig2006}, discovered two photometrically variable binary central stars (NGC~6337 and NGC~6026; Secs.~\ref{sssec:ngc6337} and \ref{sssec:ngc6026}) in an initial sample of 8 objects that were selected from a list of central stars suspected of having stellar-mass binary companions based on morphological considerations \citep{Soker1997}. Interestingly, two detections out of 8 observed targets bodes well for a correlation of certain morphological features with binarity and, by implication, for a higher binary central star fraction in general. Yet, we need to wait for more results before anything can be stated with statistical relevance.
  
\section{Summary and conclusions}
\label{sec:conclusions} 
 
Of the 12 binary central stars whose photometric variability can be ascribed to an irradiation effect (though we remind the reader that one could also be an ellipsoidal variable and that 5 need more data), 9 have a period smaller than or equal to a day. The remaining three have periods smaller than 3 days. Since the total number of objects is low, it is very difficult to draw a conclusion as to the bias of this survey, and hence whether we expect more systems at longer periods. Population synthesis models do not predict the observed decline of the number of post-CE binaries for periods longer than $\sim$0.5 days. On the other hand this decline is unlikely to be justified by a strong bias against the detection of such systems. We will have to wait for the results of our new survey to shed light on this issue. 

In two, but possibly all four of the systems for which a spectrum of the ``night" side of the secondary exists, the night side of the secondary is that of a hotter, more massive object than would be deduced from the secondary's mass. For four out of five systems for which the secondary's radius and mass have been determined, the radius is larger than for a main sequence star of the given mass (by up to a factor of two). For the fifth system, A~41, the opposite is true.  
This trend (A~41 being the exception, see below) seems to imply that the secondaries are larger and more luminous than they would be if they were main sequence stars of that mass. In the literature, the typical explanation for this effect is that the star expanded during the CE because it accreted mass and it has not yet recuperated its thermal equilibrium size. This is doubtful (though not excluded)  since the spiral-in speed during CE is typically supersonic and a bow shock therefore forms around the secondary, which may prevent material from accreting \citep{Sandquist1998}. 

Another possibility is that the star is puffed up because it is irradiated and heated. This would also explain why the spectrum of the night side tends to be that of a hotter (rather than only larger) star. \citet{Harpaz1991,Harpaz1994,Harpaz1995} study the irradiation of a main sequence star by a neutron star companion in an asynchronously rotating binary and compare the heating and expansion of the main sequence star to that of a main sequence star that is continuously and isotropically irradiated. They find that indeed the expansion is significant. It is not clear however how this conclusion could be applied equally to a star that is being irradiated on one side only (because the orbital period is synchronous with the cool star's rotation). 

The reason why the model parameters of A~41 behave opposite to those of the other systems (radius {\it smaller} than implied by the mass) might be that, as we explained in Section~\ref{sssec:a41}, in this system the photometric variability is due to an ellipsoidal variation and not irradiation, as explained by the alternative model (cf. Tables~\ref{tab:stellarparameters} and \ref{tab:stellarparameters2}). 

Another consideration regarding our understanding of irradiated atmospheres and its application to the modeling of central star light-curves are the impact azimuthal energy transport  and secondary spin synchronization with the orbital rotation \citep[][and references therein]{Cranmer1993,Brett1993}. Leakage of radiation from the ``day" side leaks into the ``night" side (for instance for those secondaries with a strong temperature gradient due small orbital separation), would reduce the contrast between day and night sides, and decrease the variability amplitude. Not including the azimuthal energy transport would therefore have repercussions on the interpretation of the light curve, leading to erroneous results. As for phase locking, one wonders whether there would be a substantial difference in the light-curve properties of phase locked and non phase-locked systems. If, for instance, systems with longer periods, hence less likely to be phase locked, had a smaller brightness difference between ``day" and ``night" sides, we might have a natural explanation of why systems with longer orbital period are harder to detect.


In the literature the consensus is almost unanimous that these systems are not CVs, in that they do not show the spectral signature of disks (broad and possibly variable emission lines). We interject, however, that optically thick disks might not exhibit broad emission lines, except at low inclinations, since the radiation only reaches us from the outer parts of the disks which are rotating slower. In addition flickering has not been observed in any of these objects. If any of these post-CE central stars had accretion disks, it would demonstrate the ability of the CE interaction to produce CVs with no need for other period-altering mechanisms. Evidence that some central stars of PN do undergo CV-type activity has recently been found in the form of a CV outburst (Nova Vul 2007 No. 1; Wesson et al., in preparation) in the middle of an old PN detected in pre-outburst images from the IPHAS survey (INT Photometric Halpha Survey; \citealt{Drew2005}). A nebula detected by IRAS around nova GK~Per \citep{Bode1987} has also been claimed to be an old PN.

In conclusion, we have demonstrated that the central star short period binary fraction of 10-15\%, is not  as free from difficulties as previously thought. The period distribution of the known systems and the maximum period detected by this survey do not tally with expectations nor can they be readily explained by likely biases. It remains true that the very small number of known binaries (including the fact that about half of them have very little information) is a hinderance when trying to understand this sample.  If we are to make any progress on the binary channel for the formation and shaping of PN, there is an urgent need to observe spectroscopically those objects that are not yet confirmed as irradiated binaries and to find more binaries to determine their characteristics as a group and to determine the binary fraction and period distribution in a more systematic way.

 \acknowledgments
OD is always grateful to Howard Bond for a continuing collaboration and the many conversations through which her work is constantly refined. We acknowledge the useful comments made to this manuscript by Howard Bond, Noam Soker, George Jacoby and Kristen Menou. This work has been in part supported by NSF grant AST-0607111 (PI: De Marco).

\appendix
\section{The known photometrically-variable binary central stars of PN}

\subsection{The eclipsing, irradiated binaries}

\subsubsection{BE~UMa}
\label{sssec:beuma}

BE~UMa has been known to be a variable since the early 70s \citep{Kurochkin1971}, but was diagnosed to be a binary in the 80s \citep{Ferguson1981}. \citet{Ando1982} presented the light-curve of BE UMa that shows a deep (but possibly not total) eclipse that lasts for approximately 2\% of the orbit and is $\sim$4~mag deep in the 3330-3700~\AA\ band and $\sim$1~mag deep in the 5265-6005~\AA\ band. The light curve also displays an irradiation effect (outside of eclipse) with an amplitude of 1.06~mag in the 3330-3700~\AA\ band and 1.4~mag in the 5265-6005~\AA\ band (similar to $u$ and $y$ in the Str\"omgren four-color system). Both the irradiation and the eclipse light-curves have a period of 2.29~days. The difference in the amplitude in these two bands is expected  and indicates that the secondary's hottest point is not extremely hot \citep[5000-8500~K;][]{Ferguson1987}, as is the case for some other systems. The amplitude of the variability is extreme, the second largest in our group (after that of K~1-2). This is likely due to the edge-on view we have of this eclipsing system and the high effective temperature of the primary. 
The system's $V$ magnitude was measured at light minimum, but out of eclipse (just before the light dimming because of the eclipse) to be 16.15 \citep{Ferguson1987}.

The emission line spectrum deriving from the heated hemisphere of the secondary star is typical of irradiated systems, with emission lines of H, HeII, carbon, nitrogen and oxygen. There is an indication that this spectrum has suffered a secular evolution \citep{Crampton1983}. These emission lines, which are not contaminated by the extremely faint PN \citep{Liebert1995}, are also relatively narrow and exclude cataclysmic variable (CV) -type activity in this system.

The masses of the primary and of secondary stars are determined by \citet{Ferguson1999} to be 0.70$\pm$0.07 and 0.36$\pm$0.07~\msun, respectively, from a fit to the radial velocities of the primary absorption lines (from HST/UV spectra) and the emission lines of the irradiated side of the secondary, which move together with the secondary's center of light (this motion was corrected to that of the center of mass). They also fit the eclipse $U$,$B$,$V$ and $R$ band photometry of \citet{Wood1995}, thus obtaining radii for the two stars (0.078$\pm$0.004 and 0.72$\pm$0.05~\rsun, respectively). The primary's radius  is consistent with the radius obtained from a spectroscopic analysis of the primary.  

As is common for this type of post-CE short period irradiated systems (see later Sections), the mass of the secondary implies a later spectral type (M3V in this case) than either the stellar radius or the spectrum of the night side of the secondary. Interestingly, the radius, 0.72~\rsun, is exactly that of a K5V star, the spectral type determined by \citet{Liebert1995} from a spectrum of the system at minimum light. The temperature determined for the secondary, 5800~K, is instead higher than either a K5 or an M3 star. It corresponds in fact to a spectral type of G2. This however is possibly the most uncertain measurement, since the various analyses present in the literature could never fully reproduce the spectral energy distribution of the system at minimum light, implying that there is a residual luminosity which is not simply that of the night side of the secondary. In conclusion, the secondary star is likely bloated, as is the case for many other irradiated systems (see below). This is attributed \citep{Ferguson1999} to the recent emergence from the CE phase, where the secondary star could have accreted matter and expanded. In Section~\ref{sec:PvA} we discuss an alternative scenario to explain these observations.

\subsubsection{A~46}
\citet{Bond1980} discovered V477~Lyr, the central star of A~46, to be a partially eclipsing binary central star of PN with a period of 11.3 hours, only the second to be discovered (after UU Sge, the central star of A~63; Section~\ref{sssec:a63}). The primary minimum, caused by a partial eclipse of the primary lasting 48 minutes, or 7\% of the phase, was measured to be 1.4 mag deep while they reported the sinusoidal modulation of the light-curve to have an amplitude of $\sim$0.5 mag and likely to be due to an irradiation effect.
\citet{Pollacco1994} presented and modeled the $V$-band photometric light curve as well as the radial velocity curves in an analysis similar to that carried out by \citet{Pollacco1993} for UU~Sge. From their Figure~6 we measured the amplitude of their light-curve, excluding the eclipses, to be 0.3~mag. The parameters they derived from the light and radial velocity curve analyses are reliant on the primary's effective temperature, which could only be estimated from a variety of indirect methods such as the He II Zanstra temperature or the ratio of  far and near UV integrated light \citep{Schoenberner1984}.
They also estimated the secondary star mass to be 0.15~\msun (or a spectral type of M6V). However the secondary's radius, 0.46~\rsun, is quite a bit closer to the radius of an M2V star, while the polar temperature of 5300~K would correspond to a primary spectral type of G0V. They therefore conjectured that the secondary has accreted mass during the CE and expanded, a similar conclusion to that reached for BE~UMa binary. We defer a discussion of this effect to Section~\ref{sec:PvA}. The minimum $V$ magnitude (but out of eclipse) of V477~Lyr was meeasured by \citet{Kiss2000} to be 15.6~mag.

\subsubsection{A~63}
\label{sssec:a63}

\citet{Bond1978} and \citet{Miller1976} determined the period of the eclipsing binary UU~Sge, in the middle of the PN A~63 to be 11.3~hours and the system to also exhibit a reflection effect. UU Sge was first discovered to be an eclipsing binary by \citet{Hoffleit1932}. Its PN was discovered by \citet{Abell1966}, but the coincidence of the binary and the PN was not realized till 1976 \citep{Bond1976}. From Figure~2 of \citet{Bond1978}, we measured the amplitude of the reflection effect (excluding the eclipse data points) to be 0.26~mag in the $B$ band. \citet{Pollacco1993} presented a $V$-band light-curve (with an amplitude of 0.36~mag) and a radial velocity curve analysis. \citet{Walton1993} reported no $B-V$ color change with orbital phase, which would indicate that the irradiated side of the companion is very hot. Indeed the analysis of \citet{Pollacco1993} determined a temperature on the bright side of the secondary of $\sim$25\,000~K. They also deduced that the primary effective temperature is $T_{\rm eff} \sim$125\,000~K from the ratio of integrated far and near UV light \citep{Schoenberner1984}. This estimate is higher than was previously published (35\,000-45\,000~K; \citealt{Bond1978,Walton1993}), but it appears to be the best value to adopt in the absence of a full spectral fitting. 

From a secondary's mass of 0.29~\msun, a spectral type of M4V\footnote{We note however that \citet{Pollacco1993} find a large discrepancy between their solution and theoretical evolutionary tracks in that the location of the primary on the Hertzprung-Russell diagram implies a primary mass which is much higher than indicated from the radial velocity solution. One way out of this discrepancy might be if the physical location of the emission line region was not that assumed by the model. This would mean that even for this eclipsing binary, the masses are ill determined.} can be derived. This is at odds with the spectral type determined by \citet{Walton1993}, G7V, from a fit of the spectrum during the eclipse of the primary (although \citet{Pollacco1993} revisited the red spectrum of A~63's core and from a temporal analysis of the radial velocity of specific diagnostic lines, cannot confirm the analysis of \citet{Walton1993}). It is also at odds with the spectral type derived from the radius (M1V) or the temperature (F1V). There are a few emission lines of NIII and CIII-IV in the 4650-\AA\ region of the spectrum from irradiation (they move in anti-phase to the absorption lines), but the irradiation spectrum in this object is not as rich as that exhibited by V~644~Cas. The inclination of the orbit is found to be 87.7~deg, very similar to that found by \citet{Mitchell2007} purely from a kinematical analysis of the PN. \citet{Mitchell2007} also characterized two jets on either side of the star. The very similar inclination angles determined by the stellar light-curves and the PN kinematical analysis solidly tie the morphology of this object to its binary central star. The minimum $V$ magnitude of UU~Sge was measured by \citet{Bell1994} to be 19.20$\pm$0.07~mag.

\subsection{The non-eclipsing, irradiated binaries}

\subsubsection{HFG~1}
\label{sssec:HFG1}

\citet{Grauer1987} presented a photometric light curve of V664~Cas, the central star of the PN HFG~1, that has a period of 14.0 hours and an amplitude of 1.1 mag in the $B$ band that is due to irradiation effects from this close binary.  \citet{Acker1990} presented a spectrum of the central star which has strong H and He emission lines. They concluded this object to have an accretion disk, as is the case for polar CVs. However there is no evidence of fast variability of the light curves (``flickering") as is expected in the case of accretion disks and the spectrum can be explained with a primary and an irradiated secondary \citep{Shimanskii2004,Exter2005}. 

\citet{Shimanskii2004} presented an analysis of the light and radial velocity curves. From an analysis of the light curves in the $U$, $B$, $V$, $R$ and $I$ bands they refine the period and determine that there is a slight effect due to ellipsoidal variation with an amplitude which is only 2\% of the amplitude due to the irradiation effect. The similarity of the photometric variation amplitude in all bands is interpreted as a hot spot on the surface of the secondary observable at all projections, dominating the brightness and color of the system. This implies that the orbital inclination is small (28$\pm$2). From the radial velocity curve of the emission lines produced on the irradiated side of the secondary, they determined that the orbit has a non-zero eccentricity (0.09$\pm$0.01). They however favored a solution with zero eccentricity, despite it not fitting the radial velocity curve as well, since no other system in this class has ever been observed to have a non-zero eccentricity. A non-zero eccentricity is however not excluded for post-CE systems on theoretical grounds \citep{DeMarco2003}. 

The secondary radial velocity amplitude they derive is converted to the center of mass radial velocity amplitude by assuming a specific stellar rotation ($V \sin i = 30$~\kms). By assuming a range of plausible masses for the primary (0.55-0.60~\msun) and assuming a range for the mass function determined from the irradiation lines and a range of inclinations (26-30~deg) they determined a range of secondary masses of 0.90 -- 1.30~\msun. As is the case for K~1-2 (Sec.~\ref{sssec:k1-2}), the secondary in this system is likely to be more massive than the primary. 

\citet{Shimanskii2004} also modeled the light-curve and the best parameters derived are those listed in Table~\ref{tab:stellarparameters}. We note, however, that the primary's temperature is a free parameter and higher than what was indicated by \citet[50\,000-60\,000~K]{Heckathorn1985}, for whose range of temperatures the model amplitudes were too small. We wonder whether this is a similar effect to that found by \citet[Section~\ref{sssec:k1-2}]{Exter2003b} who reverted to a higher than possible irradiation efficiency to model the very large irradiation amplitude. The large secondary radius must be approaching the Roche Lobe radius and this would be in line with the small amount of ellipsoidal variation. If so then there might be some accretion of secondary material onto the primary. However, it is pointed out that this system cannot have an appreciable accretion disk around the primary, or this would be obvious as an enhancement of the emission lines at minimum brightness - a thing which is not observed. Finally, it is possible that the CNO abundances on the secondary might be super-solar, denoting accretion of primary material during the CE phase.

\citet{Exter2005} presented another analysis of this system, demonstrating however the impossibility of pinning down its basic parameters. $K_2$, the radial velocity semi amplitude of the secondary, determined from the emission lines associated with the secondary, is measured to be 68$\pm$6~\kms, but $K_1$ could only be pinned down to be in the range 49-89~\kms. If the lower value were adopted, and a primary's mass of 0.63~\msun\ were chosen from their Table~2 (this being close to the mean value for central star masses), then a match to the data would be achieved with an inclination of 29~$\deg$, and a secondary mass of 0.41~\msun\ different from the model of \citet{Shimanskii2004} who favored a scenario where the secondary mass is actually larger than the primary mass by a factor of 1.6. Because of this difference we report both these models in Table~\ref{tab:stellarparameters}. \citet{Exter2005} also determined a spectral type from the night side of the spectrum ranging between F5V and K0V and a sub-stellar temperature on the secondary surface of 22\,000~K. 
 
\subsubsection{K~1-2}
\label{sssec:k1-2}

\citet{Bond1979} discovered that VW~Pyx, the central star of the PN K~1-2, is a periodic photometric variable with an amplitude of nearly 1.4 mag (between 16.4 and 17.8 magnitudes in the ``blue") and a period of 16.1 hours (revised to 16.2 by \citet{Exter2003}). They ascribed the variability to pulsations, though \citet{Bond1987} later suggested that the variability is instead due to the binary nature of the star. \citet{Kohoutek1982b} determined the Zanstra (HeII) temperature of the star (79\,000-91\,000~K). \citet{Exter2003} reported that the central star of this post-CE binary is a single-lined or possibly a double-lined spectroscopic binary, where the primary star absorption lines can barely be resolved, while the emission line spectrum of the irradiated secondary dominates. The spectrum is very similar to that of V664~Cas (HFG~1). They also showed a $V$-band photometric light-curve with an amplitude of 1.35~mag and say that their $B$ and $V$ band brightness are the same within 0.03~mag. Once again this implies a very hot temperature of the irradiated face of the secondary star -- they determined a sub primary star temperature of 40\,000~K. Their radial and light-curve analysis implied that the secondary's mass must be larger than the primary's mass for any reasonable assumption. The inclination is not constrained and is assumed by them to be 50~deg (which is similar to the value of $\sim$40~deg determined by \citet{Corradi1999} from an analysis of the PN's jets). If the primary mass is assumed to be 0.6~\msun, a reasonable value for a post-AGB star, a secondary mass of $>$0.74~\msun\ is deduced, where the greater-than sign is due to the fact that the secondary's radial velocity amplitude is derived from the emission lines which sample the center of light and not the center of mass of the secondary. There is a small (4\%) ellipsoidal variation. Most interesting is that there is no way of modeling the light-curve without assuming an unrealistically high heating efficiency. This seems to go hand in hand with the large amplitude of the irradiation effect, the largest of the entire group and much larger than for any of the four known eclipsing binaries.  

\subsubsection{DS~1}

KV Vel, the central star of PN DS~1 was discovered to be a variable by \citet{Drilling1984}, with a period of 8.5 hours. \citet{Drilling1985} showed this star to be a double-lined spectroscopic binary and determined the primary stellar temperature to be 77\,000$\pm$8000~K from the ratio of the integrated near to far UV flux from an IUE spectrum \citep{Schoenberner1984}. \citet{Kilkenny1988}  presented a light-curve in the $U,B,V$ and $I$ bands, which confirmed the amplitude of 0.6 mag. They measured the central binary brightness to vary between $V$=11.75 and 12.30~mag and detected a color variability indicating that the system is more variable at $I$ than $U$, likely due to the fact that the hottest point on the secondary is not particularly hot.

\citet{Hilditch1996} presented a photometric and spectroscopic study of this object. The photometric variability is sinusoidal and best explained as due to an irradiation effect. The amplitude of the irradiation variability is slightly larger in $I$ than in $B$ ($\Delta B = 0.51$~mag $\Delta V = 0.55$~mag $\Delta I = 0.58$~mag). 
No hint of an eclipse was seen. This fact, and the irradiation amplitude constrained a combination of inclination angle,  secondary radius, limb-darkening coefficient and irradiation efficiency. The inclination angle is constrained to the range $50^o < i < 70^o$, within which \citet{Hilditch1996} adopted a value of $i = 62.5^o \pm 1.5^o$. The stellar mass ratio is known from the amplitudes of the RV curves determined from the primary absorption lines and emission (irradiation) lines associated with the secondary ($M_2/M_1$=0.36). From these constraints, and a primary stellar mass assumption of 0.63~\msun, a secondary mass of 0.23~\msun\ was determined. From this we would conclude that the secondary has spectral type M5V. Its radius, however, was determined to be 0.402~\rsun, corresponding to an M3V star. Once again we see evidence that the main sequence primary is larger than implied by its mass. With a primary radius of 0.157~\rsun\ and a luminosity of 776~\lsun, the primary would have a mass smaller than 0.57~\msun\ (from a comparison with the hydrogen-burning tracks of \citet{Vassiliadis1994}) and somewhat inconsistent with the mass determination (0.63~\msun) of \citet{Hilditch1996}. 
 
\subsubsection{Sp~1}
\citet{Mendez1988a} suggested that the unusual spectrum of the central star (ESO225-2) of PN Sp~1, which exhibits high temperature absorption and emission lines as well as lower temperature emission lines, could be explained if it is assumed to be a close binary system that exhibits irradiation effects. This object was confirmed to be a binary by \citet{Bond1990}, having a period of 2.9 days, relatively low amplitude photometric variations of $\sim$0.1 mag in the $B$ band (no light-curve has ever been published), and no eclipse. This, they supposed, could be due to the fact that we are observing the system nearly pole-on, in line with the PN's ring morphology, if we interpret it as a bipolar PN seen pole-on.

\subsubsection{NGC~6337}
\label{sssec:ngc6337}

\citet{Hillwig2006} presented a  photometric light curve of ESO~333-5, the central star of PN NGC~6337 in the $B$, $V$ and $R$ bands.  The sinusoidal modulation has a period of 4.2~hours and an amplitude of 0.41~mag in the $V$ and $R$ bands and 0.51 in the $B$ band.  They suggested that the orbital period is the same as the photometric period and it is probably due to an irradiation effect.  They could model the light curve with several sets of parameters, of which they presented two sets for the two extreme assumptions of primary temperature. A primary mass of 0.6~\msun\ was assumed for the primary and an upper limit for the mass of the companion of 0.35~\msun\ was used based on a main sequence radius which would fit inside the orbit of this very close binary (although it is possible that this upper limit is too high, since every time masses and radii for secondaries in irradiated systems are determined, the radii are always larger than those of main sequence stars of the given mass; if this were the case for this binary, then the secondary mass would be smaller). For this mass, a relatively cool primary star temperature has to be selected to fit the data (45\,000~K). A higher temperature value (105\,000~K) with different model parameters, is however a better match to the data.

\subsubsection{Hf~2-2}
\label{sssec:hf2-2}

The central star ESO~457-16 (Wray~16-414)  of the PN Hf~2-2 was discovered to be a photometric variable by \citet{Lutz1998} by inspecting MACHO data (MACHO ID 139.32591.221). They determined a period of 9.6 hours. A hard copy of its unpublished light-curve folded in phase space, was kindly made available by Howard Bond (priv. comm.) and measured by one of us (OD) to have a photometric amplitude in $V$ of 0.325~mag and a range between 17.07 and 17.40~mag. The scatter of datapoints is about 0.1~mag, although the error bars on each data point span $\sim$0.02~mag. \citet{Liu2006} determined the spectro-photometric $B$ and $V$ brightness for this object to be 17.04 and 17.37~mag, respectively, indicating that they measured the spectrum near minimum light. They do not report the detection of an irradiation spectrum, possibly because they observed the system near minimum light. \citet{Perek1967} cited $B = 18.0$~mag. Inexplicably, this $B$-band magnitude is a full magnitude dimmer than the measurement of \citet{Liu2006}, whom we concluded to have measured the system at minimum light. If $B-V$=--0.33 \citep{Liu2006} then the $V$-band magnitude of \citet{Perek1967} would be 18.33~mag, 1.26~mag fainter than the MACHO minimum light observation and inconsistent with a light variability range of 0.33~mag determined from the MACHO data. This would imply a $B$ band variability of $\sim$1~mag, which is suspiciously much larger than the 0.325~mag measured from the $V$ light-curve. They determined a Zanstra (HeII) temperature in the range 50\,000--67\,000~K and fitted a 67\,000-K blackbody to the stellar spectral energy distribution, reaching a satisfactory fit.

\subsubsection{A~65}
\citet{Bond1990} listed the central star of A~65, ESO~526-3, to have a light variability with a period of approximately 1~day and an amplitude of approximately 0.5~mag in the $V$ band, from an incomplete and unpublished light-curve. \citet{Walsh1996} later studied the central star and nebula, measuring $V = 15.4$~mag. They note that the \citet{Abell1966} photometry is 0.5~mag fainter than their own, although his colors agree with their determination. This might be due to the fact that \citet{Abell1966} measured the central star at minimum light, so that $V$=15.4~mag could be the maximum light magnitude. \citet{Walsh1996} call A~65's central star an eclipsing binary. However, there is no evidence in the literature that the 0.5~mag declines are indeed due to an eclipse and not to an irradiation effect. We therefore list it simply as a photometric variable. \citet{Walsh1996} measured the spectrum of the central star and concluded it is very similar to that of a CV (and to the spectrum of HFG~1; \citealt{Acker1990}), in that it has strong H and He lines in emission. However this spectrum does {\it not} present the broad emission line wings characteristic of CV spectra that are due to fast rotating gas in a Keplerian disk around the hot primary and there is no sign of flickering, which would indicate the presence of an accretion disk even in the absence of broad lines. It is more likely that this spectrum is once again due to the irradiated surface of the companion.

\subsubsection{A~41}
\label{sssec:a41}
\citet{Grauer1983} reported that MT~Ser, the central star of A~41, has a $U$ and $B$-band photometric variability of 0.15~mag with a period of 2.7 hours that they ascribed to an irradiation effect, although they could not exclude that it could also be produced by ellipsoidal variability, in which case the orbital period would be twice as long.  

\citet{Green1984} reported results obtained from their higher resolution spectroscopy where they detected the absorption line spectrum of the hot central star. From the ratio of He I and He II lines they derived an effective temperature for the primary of 50\,000$\pm$5000~K in agreement with an upper limit from a Zanstra method determination. From their spectral analysis and an assumed primary mass of 0.6~\msun, they derived a primary radius of 0.13~\rsun. They also pointed out that the secondary star in this system could be an evolved subdwarf (though not a WD since such a small secondary would not show irradiation effects) instead of a cool main sequence star. They also determined a substellar temperature of 16\,000~K and a temperature of the secondary, in case it is a main sequence star, of 3000~K. They noted that the fact that the $U$ and $B$ light-curves look very similar implies that the U flux is enhanced possibly from Balmer emission. They did not mention whether an irradiation emission line spectrum was present in their observations, and from their figures there does not appear to be one.

\citet{Bruch2001} reanalyzed the $B$-band light-curve and could not distinguish between two possible models, the first containing a cool companion irradiated by the hot central star as well as a small ellipsoidal effect, and the second where the companion is an evolved hot star and the variability is entirely due to ellipsoidal variability. Without a radial velocity curve there is no way to distinguish between these two models. We therefore report the parameters of both these models in Tables~\ref{tab:stellarparameters} (their model 1.1) and \ref{tab:stellarparameters2} (their models 2.2.1 and 2.2.2 ) and use their model 1.1, that of an irradiated secondary with a small ellipsoidal variability component, to carry out our analysis. This model fits their $B$-band light-curve, which we measured (using their Figure 1) to have an amplitude of 0.255~mag.  

Finally we note that although we include this star in the irradiated binaries, the absence of an irradiation spectrum argues against the photometric modulation being due to an irradiation effect. We  also noticed that if we adopt the irradiation model from Table~\ref{tab:stellarparameters}, the secondary star spectral type determined by assuming that the derived mass is that of a main sequence star (K9V) is {\it earlier} than the spectral type derived by assuming that the {\it radius} is that of a main sequence star (M1V). This would imply that the secondary is smaller than for a main sequence star of that mass. This is opposite to the five other systems where this comparison can be carried out (see Table~\ref{tab:stellarparameters}) and might indicate that indeed the irradiation model is not correct for this objet and that we should therefore adopt the ellipsoidal variability model from Table~\ref{tab:stellarparameters2}.
 

\subsubsection{HaTr~4}
\citet{Bond1990} report a $V$-band photometric variability of 0.4~mag with a period of 1.71~days and state that the  light curve was fragmented but appeared to be sinusoidal (no light-curve has ever been published). Very little is known about this object, so we have to treat it with caution until we know whether the variability is indeed due to reflection  and not to ellipsoidal variation.

\subsection{Other photmetrically-variable objects}
\label{ssec:other}

There are five central stars that have been discovered because of, or that are known to have, photometric variability, but that are clearly {\it not} irradiated binaries. In fact one, and possibly two, of these five central stars might not even be a binary. We list them below for completeness.

\subsubsection{NGC~2346: an A5V single lined spectroscopic binary orbited by a dust cloud}
\label{sssec:ngc2346}

V651~Mon, the central star of the bipolar PN NGC~2346 has an A5V spectral type \citep{Mendez1978} and is too cool to ionize the PN, from which it can be deduced that that it has an unseen hot companion.  \citet{Mendez1981} reported the central star to be a single-lined spectroscopic binary with a period of 15.99 days. \citet{Kohoutek1982a} found the central star to be a partly-eclipsing binary. However, the light-curve of the system was monitored during many years and ample evidence accumulated that the eclipses are due to a dusty cloud, not another star \citep{Costero1986}. This dusty cloud appears to be orbiting with a (variable) frequency, related, though in a non-trivial way, to the binary period  \citep{Costero1986}. \citet{Kato2001} reported a new feature of the light-curve, consisting of a prolonged minimum ($\sim$400~days in 1996-1997). In summary it is probable that a former disk, possibly ejected during the CE that formed this PN, is now breaking up, leaving behind dusty fragments, one or more of which pass periodically in front of the star \citep{Mendez1981}. There is no indication of an irradiation effect in this already complex variability history.

\subsubsection{Hb~12: a cool variable central star.}
\label{sssec:hb12}
 
\citet{Hsia2006} observed NSV~26083, the central star of the well-studied PN Hb~12, to have a photometric variability in the $R$ and $I$ bands, with amplitudes of 0.06 and 0.08~mag, respectively, with a period of 3.4~hours. They interpreted this effect to be due to the eclipsing of an cool companion by the hotter central star. The fact that the light-curve dips ascribed to the companion eclipsing the primary are much shallower in $R$ than they are in $I$, is explained by the authors as H$\alpha$ emission from the irradiated secondary being included in the $R$ band. This would be strange since, during the eclipse,  the ``night", non-irradiated side of the secondary would be in view. However, contamination from the bright PN would contribute to the reduction of the putative eclipse depth. \citet{Hsia2006} also claimed the detection of spectral absorption features consistent with a spectral type of G to early K. Assuming the primary effective temperature and mass of 31\,800~K and 0.8~\msun, respectively (determined by \citet{Zhang1993} from an analysis of the nebula), they derived $M_2$$<$0.443~\msun, (corresponding to a spectral type later than M1V). However, to determine this secondary mass limit, they assumed a specific mass ratio and a main sequence mass-radius relation which, we know from several other irradiated systems, not to apply. Although a few emission lines are seen in the spectra presented by \citet{Hsia2006} it is not clear from their figures or their analysis whether any of the features can be ascribed to an irradiation effect. 

Looking at their $I$ band photometry (their Fig.~2) there seem to be two modulations, one with a smooth sinusoidal shape and one consisting of dips lasting about 40~min or $\sim$20\% of the phase. Strangely, the ``dips" appear near the maximum brightness of the light-curve plotted as a function of time (their Fig.~2), but appear near the {\it trough} of the light-curve plotted as a function of phase (their Fig.~4; as is expected for an eclipsing, irradiated binary system -- see for instance the light-curve of A~63 in \citealt{Bond1978}). Strangely, the data scatter is smaller when plotted as a function of time, but increases in the data folded into phase, an opposite behavior to what is expected for truly periodically variable data. In addition the three observed eclipses strangely have non-constant and non-symmetric shapes and do not resemble any of the eclipse lightcurves observed in other systems. The sinusoidal variability in the $I$ band, where the dips can clearly be distinguished, is no larger than a couple of tens of a magnitude, slightly smaller than the absolute depth of the eclipse ``dips". Finally, such a short period implies a separation of a few solar radii, hardly large enough to accommodate the central star of Hb~12 which, being relatively cool, must have a radius of the same order of magnitude. All these considerations make us wonder what the real nature of the variability of Hb~12 might be, and whether the variability might not be more akin to orbiting dust as is the case of NGC~2346 (Sec.~\ref{sssec:ngc2346}).
In conclusion this system is in need of additional observations before it can be confirmed as a binary. If indeed the sinusoidal modulation is due to irradiation then the small amplitude could be due to the particularly low temperature of the primary.

\subsubsection{SuWt~2: a peculiar A+A binary}
\label{sssec:suwt2}

\citet{Bond2000} listed a photometric period of 4.9~days for the central star (GSC2 S2122230582) of the PN SuWt~2.  Unfortunately he did not include information about the band or the amplitude of the variation. \citet{Bond2002} say that this binary is eclipsing, and its inclination ($\sim$90~deg) is therefore similar to the inclination of the ring-shaped PN around it \citep[$\sim$64~deg;][]{Smith2007}. \citet{Exter2003b} classified this object as a double-lined spectroscopic and eclipsing binary for which {\it both} stars appeared to be of spectral type A. They therefore conjectured this to be a triple system, where the actual (hot) central star is as yet to be found. Interestingly, they found no trace of a hot component in the IUE spectra. They also presented the radial velocity curves of the system and determined the spectroscopic period to be 4.9 days. If there is a third hot star in the system that is responsible for the PN, it should be a distant companion since close triple systems are not stable. If so, it is difficult to understand how the PN can have been shaped by the interaction \citep{Smith2007}. 

\subsubsection{Sh~2-71: a mis-identified central star?}
\citet{Kohoutek1979} identified the central star of PN Sh~2-71\footnote{Although this object has 18 names from various catalogue entries, there is no single name that meaningfully identifies the central star.} to be the bright ($V \sim 13.5$~mag) object slightly east of the PN's center, and measured this object to have a light variation with an amplitude of $\sim$0.7~mag.  \citet{Feibelman1999} detected a variable Mg~II line at 2800~\AA\ and concluded that the variability is either caused by the orbiting hot companion or by star spots on the cooler star. However, Frew (2008) \nocite{Frew2008} noticed a hot star in the {\it geometric center} of the PN with a magnitude consistent with it being the central star (considering published values for the distance and reddening to this PN). If this object is truly the central star, the star analyzed by \citet{Kohoutek1979} could be a very peculiar and rare chance alignment of a PN and a short period evolved binary. Alternatively, the hot central star and the peculiar binary are associated. Until more data is gathered, a verdict cannot be reached and we cannot confirm this object as a bona fide binary central star of PN.

\subsubsection{PN G135.9+55.9: a double degenerate system}

\citet{Napiwotzki2005} presented new observations, which confirm the central star (SBSS1150-599) of PN G135.9+55.9 to be a single-lined spectroscopic binary with three possible periods: 0.94, 1.5 and 3.6~hrs (see also \citet{Tovmassian2004}). They also detected a photometric variability with a strong periodicity at 3.9~hrs (similar to the longest of the three spectroscopic periods) and a 0.14~mag amplitude (where we measured the amplitude for the ``blue" light-curve presented in their Fig~4). They convincingly demonstrated that this is not due to  irradiation of a main sequence companion by a hot primary. They instead interpreted the light-curve as due to an ellipsoidal variation {\it of the primary central star} as the gravity of the secondary (unseen but most likely evolved) companion makes the central star partly fill its Roche Lobe. This is a very different situation from most of the central stars in this paper, where the photometric variability, whether due to irradiation or ellipsoidal variation, is due to the companion. In the light-curve of PN~G135.9+55.9, however, alternate minima are much brighter. \citet{Napiwotzki2005} interpreted this observation as extra light originating on the central star because of being irradiated by an unseen {\it neutron star} companion. Alternatively, they suggested that the extra light could derive from a tilted disk around the compact companion, fed by the central star. 

The gravity and effective temperature of the primary place it on the stellar evolutionary track of a 0.88~\msun\ object, but its location in the Galactic halo, as well as its low atmospheric metallicty, kinematic age and low metallicity of the PN, convincingly indicate it to be an old, lower mass object \citep[0.55~\msun;][]{Tovmassian2004}. It is not likely that this star is out of thermal equilibrium, since the relaxation timescales are rather short. \citet{Tovmassian2004} suggest this object to be a post-helim shell flash star (a ``born-again" star -- \citet{Iben1983}), although its photospheric abundances are not hydrogen-defficient, as should be the case for those objects \citep{Werner2006}.  

This is a very interesting binary since its total mass is very close to the Chandrasekhar limit and the orbital separation is such that the system could merge in 1~Gyr, making it one of a few potential Type Ia supernova progenitors \citep{Napiwotzki2003}.    

\subsubsection{NGC~6026: a central star with a WD or a sub-dwarf companion}
\label{sssec:ngc6026}

From the radial velocity curve, \citet{Hillwig2006} determined an orbital period of 12.7~hours for ESO~398-7 (Wray 16-201), the central star of the PN NGC~6026. This is twice the period deduced from the light-curve which means that the 0.19~mag-light modulation is not due to irradiation effects, but to ellipsoidal variation of the primary which is distorted by gravity of the secondary.  No irradiation is detected in this close binary, likely indicating that the companion is not a cool star, but rather WD or a sub-dwarf (horizontal branch star).  This object also exhibits an eclipse at phase 0.5 but not at phase zero. The data in hand can only be modeled with a small, very hot companion. The eccentricity of the orbit is determined to be 0.025$\pm$0.002. Because the photometric variability is not from an irradiation effect but instead from ellipsoidal variation, we have excluded this object from our analysis.

\bibliographystyle{apj}                       
\bibliography{DeMarco}

 \begin{table}\def~{\hphantom{0}}
  \begin{center}
  \caption{Close binary central stars displaying irradiation effects. I: irradiation variability; S1: single-lined spectroscopic binary; S2: double-lined spectroscopic binary (the radial velocity of the secondary is inferred from the emission lines produced on the secondary's irradiated side); Ec: eclipsing binary; El: ellipsoidal variability.}
  \label{tab:BCS}
  \begin{tabular}{llclllll}\hline
PN      &Star& $V^1$& Binary   & Period$^2$  & Var. & Band  & Irr.    \\ 
name    &name               & (mag)&  Class  &(days)    & Amp. &       &Spec.               \\
 \hline
 \multicolumn{8}{c}{Irradiated eclipsing binaries}\\ 
PNG145$^3$&BE~UMa   	&	16.15	        &	S2,Ec,I 	&	2.29	   &  1.06,1.4 & $\sim u,y$&y \\
A~46	&  V477 Lyr      & 15.6	&	S2,Ec$^4$,I	&	0.47	  & 0.3$^5$ 	&$V$&y\\
  A~63	  &UU Sge      &19.20	&	Ec,I	&       0.46	 	&  0.26$^5$,0.36$^5$ & $B$,$V$&weak\\
\multicolumn{8 }{c}{Irradiated non-eclipsing binaries}\\ 
HFG~1        &V664~Cas&	(13.4)  & S2,I,El	&	0.58       &  1.03,1.12,1.15 & $U,B,V$ &y\\
&&&&&1.15,1.12&$R,I$&\\
K~1--2	& VW~Pyx   &	(16.6)	   &	S,I	&	0.68	&  1.35        &	$V$ &y \\
DS~1	  &KV~Vel  &	12.30 &S2,I	&	0.36  &  0.51,0.55,0.58     &$B,V,I$&y\\
Sp~1    		&ESO225-2&	(14.0)	&	I	&	2.91	       	&  0.1 &$B$&y\\
N6337	&ESO333-5& 14.9	&	I	&	0.17	 	&  0.51,0.41,0.41 & 	$B,V,R$ &y?	\\

 Hf~2--2		&ESO457-16&17.37	&	I?	&	0.40		 	&  0.325 	&$V$&n?$^6$\\
 A~65	         &ESO526-3&	15.4$^7$ &	I	&	$\sim$1	         		  &  $\sim$0.5	&$ V$&y\\
 A~41$^8$	         &MT~Ser&	(16.5)	&	I?	       &	0.11         &  0.255 	&$B$&n?\\
 HaTr~4		&GSC2 S230$^9$&	17.1	&	I?	&	1.74	          &  0.4 &$V$&--\\
  \hline
 \multicolumn{8}{l}{$^1$Magnitudes from the Simbad database are in brackets and should be considered indicative.}\\ 
 \multicolumn{8}{l}{ \ \ Other magnitudes are light minimum magnitudes - see references in the text.}\\
  \multicolumn{8}{l}{$^2$Periods are here approximated to two decimal places, but are in most cases known to a}\\ 
 \multicolumn{8}{l}{\ \ much higher degree of accuracy.}\\
 \multicolumn{8}{l}{$^3$PN G144.8+65.8}\\
  \multicolumn{8}{l}{$^4$Partly eclipsing.} \\
 \multicolumn{8}{l}{$^5$These amplitudes should be 5-10\% larger due to the fact that the trough of the irradiation}\\ 
  \multicolumn{8}{l}{\ \ curve could not be accurately measured since it had to be interpolated across the eclipse dip.}\\
 \multicolumn{8}{l}{$^6$System observed at light minimum when irradiated side is partly out of view.}\\
  \multicolumn{8}{l}{$^7$Measured at light maximum.}\\
   \multicolumn{8}{l}{$^8$MT~Ser's variability can also be modeled with a hot evolved companion, in which}\\
   \multicolumn{8}{l}{ \ \     case the period would be 0.23 days and the irradiation amplitude might be very }\\
    \multicolumn{8}{l}{\ \ small or zero - see Tables~\ref{tab:stellarparameters} and \ref{tab:stellarparameters2}. }\\
   \multicolumn{8}{l}{$^9$GSC2 S2300200605}
  \end{tabular}

  %
\end{center}
\end{table}

\begin{table}\def~{\hphantom{0}}
  \begin{center}\begin{tiny}
  \caption{Close binary central stars displaying irradiation effects. Derived stellar and system parameters. Values in italics are assumed.}
  \label{tab:stellarparameters}
  \begin{tabular}{lllllllccc}\hline
      PN name     &   $M_1$ & $M_2$ & $T_1$   & $T_2$  &$R_2$& $i$ & \multicolumn{3}{c}{Spectral Type from:}    \\
                         &  (\msun)   &(\msun)       & (kK) & (K)  &(\rsun)& (deg) & Spec.        & $M_2$& $R_2$ \\
                         \hline
 \multicolumn{9}{c}{Irradiated, eclipsing binaries}\\
BE~UMa          &0.70$\pm$0.07&0.36$\pm$0.07&105\,000$\pm$5000&5800$\pm$300&0.72$\pm$0.06&84$\pm$1&K5V&M3V&K5V\\
 A~46	    &0.51$\pm$0.07 & 0.15$\pm$0.02 &60\,000$\pm$10\,000 & 5300$\pm$500 &0.46$\pm$0.03 & 80.5$\pm$0.2 &--&M6V&M2V \\
 A~63	    &0.63$\pm$0.06&0.29$\pm$0.04&$\sim$117\,500$\pm$12\,500&7300$\pm$250&0.53$\pm$0.02&88&G7V$^1$&M4V&M1V\\
\multicolumn{9}{c}{Irradiated, non-eclipsing binaries}\\
HFG~1$^2$		  &{\it 0.57}&1.09&83\,000$\pm$6000 &5400$\pm$500&1.30$\pm$0.08&28$\pm$2 &--& F9V&F5V \\
--      		  &{\it 0.63}&0.41&83\,000$\pm$6000 &-- &--&29 &F5-K0V& M2V&-- \\
K~1-2	          &{\it 0.6}&$>$0.74&$\sim$85\,000&--&--&{\it 50}&--&earlier than K2V&--\\
DS~1	           &{\it 0.63$\pm$0.03}&0.23$\pm$0.01&77\,000$\pm$8000&3400$\pm$1000&0.402$\pm$0.005&62.5$\pm$1.5&--&M5V&M3V\\

Sp~1    		&--&--&--&--&--&--&--&--&--\\
NGC~6337	&{\it 0.6}& 0.35 or 0.20&45\,000 or 105\,000&5500 or 2300 &0.42 or 0.34&28 or 9 &--&M5V or M3V&--\\

Hf~2-2		&--&--&$\sim$67\,000&--&--&--&--&--&--\\
A~65	          &--&--&$\sim$80\,000&--&--&--&--&--&--\\

A~41$^3$	        &{\it 0.6}&0.55&50\,000$\pm$5000&7517$\pm$2065&0.44&42.5$\pm$1.7&--&K9V&M1V\\
HaTr~4		&--&--&--&--&--&--&--&--&--\\

  \hline
  \multicolumn{10}{l}{$^1$From the analysis of \citet{Walton1993} disputed by \citet{Pollacco1993}.}\\
   \multicolumn{10}{l}{$^2$The first model is by \citet{Shimanskii2004} and is one of several combinations that fit the data. The second alternative model is by }\\  \multicolumn{10}{l}{\citet{Exter2003b} and is one of several possible ones, including one quite similar to that of \citet{Shimanskii2004}}. \\
  \multicolumn{10}{l}{$^3$A41's central star light-curve can also be modeled with a hot, evolved companion - see Table~\ref{tab:stellarparameters2}.}
  \end{tabular}

\end{tiny} \end{center}
\end{table}

\begin{table}\def~{\hphantom{0}}
  \begin{center}\begin{tiny}
  \caption{Photometrically variable binary central stars {\it not} from irradiation effects.}
  \label{tab:stellarparameters2}
  \begin{tabular}{llllllllll}\hline
      PN name     &P&   $M_1$ & $M_2$ & $T_1$   & $T_2$  &$R_2$& $i$ & Spec. Type & Irr. Emiss.  \\
                         &(days)&  (\msun)   &(\msun)       & (kK) & (K)  &(\rsun)& (deg) & Sec.        & Lines?\\
                         \hline
 SuWt~2		&4.9&$\sim$2.5&$\sim$2.5&--&--&--&$\sim$90&A&--\\
 NGC~2346$^1$		&15.99&1.8$\pm$0.3&$<$0.45&8000&--&--&$>$50&A5V&--\\
 Hb~12$^2$	&0.14&{\it 0.8}&$<$0.443&{\it 31\,800}&--&--&$\sim$90&G-K&?\\
 PNG136$^3$     &0.16&0.55&$>$0.82&120\,000&--&--&--&WD/NS&n\\
A~41$^4$    	        &0.23 &{\it 0.6}&{\it 0.6}&50\,000$\pm$5000&46\,000$\pm$4000&0.75&66$\pm$2&sdB&--\\
	      	        &0.23&{\it 0.6}&{\it 0.30}&50\,000$\pm$5000&46\,000$\pm$4000&0.58&67$\pm$2&sdB&--\\
  NGC~6026      &0.58 & 0.53$\pm$0.01&0.53$\pm$0.01&36\,000&134\,000$\pm$5000&0.053$\pm$0.005&82$\pm$5&WD/sdO&n\\
  \hline
 \multicolumn{10}{l}{$^1$Here the primary is the cool component, responsible for the A5V spectral type, while the secondary is assumed to be a hot central star.}\\
 \multicolumn{10}{l}{$^2$This central star's binarity needs to be confirmed.}\\
 \multicolumn{10}{l}{$^3$PN G135.9+55.9.}\\
  \multicolumn{10}{l}{$^4$The central star of A~41 can be modeled equally well with a cool companion - see Table~\ref{tab:stellarparameters}.}
  \end{tabular}
   
 \end{tiny} \end{center}
\end{table}

 \begin{table} \begin{center}
\begin{tabular}{lcc}
\hline
&Primary & Secondary\\
\hline
$M$ (\msun) & 0.6 & 0.2 \\
$R$ (\rsun) & 0.33 & 0.6 \\
$T_{eff}$ (K) & 85\,000 & 3500 \\
$L$ (\lsun) &5030 &0.05 \\
$P$ (days) &\multicolumn{2}{c}{1.6}\\
$a$ (\rsun)& \multicolumn{2}{c}{5.3}\\
$i$ (deg) & \multicolumn{2}{c}{70}\\
\hline
\end{tabular} \end{center}
\caption{Stellar and system parameters of a hypothetical system used to study the light-curve behavior as a function of different parameters.}
\label{tab:hypothetical}
\end{table}

\end{document}